\documentclass[journal,twocolumn]{IEEEtran}

\usepackage{amsmath,amsthm,amssymb,scrextend}
\usepackage{fancyhdr}
\usepackage{relsize}
\usepackage{epstopdf} 
\usepackage{floatrow}
\usepackage[font={small}]{caption}
\usepackage{multirow}
\usepackage{balance}
\usepackage[numbers,sort&compress,square]{natbib}
\usepackage{subfigure}
\usepackage{ifpdf}
\ifCLASSINFOpdf
  \usepackage[pdftex]{graphicx}
   \graphicspath{{./Figures/pdf/}}
   \DeclareGraphicsExtensions{.pdf}
\else
  \usepackage[dvips]{graphicx}
   \graphicspath{{./Figures/eps/}}
   \DeclareGraphicsExtensions{.eps}
\fi

\begin{document}

\pagenumbering{gobble} 

\title{The Potential Gains of Macrodiversity in mmWave Cellular Networks with Correlated Blocking \vspace{0.1cm}}
\author{
\IEEEauthorblockN{Enass Hriba and 
Matthew C. Valenti
 } \\
West Virginia University, Morgantown, WV, USA. \\
\vspace{-0.2cm}
}

\maketitle
\thispagestyle{empty}


\begin{abstract}

At millimeter wave (mmWave) frequencies, signals are prone to blocking by objects in the environment, which causes paths to go from line-of-sight (LOS) to non-LOS (NLOS). We consider macrodiversity as a strategy to improve the performance of mmWave cellular systems, where the user attempts to connect with two or more base stations. An accurate analysis of macrodiversity must account for the possibility of correlated blocking, which occurs when a single blockage simultaneously blocks the paths to two base stations. In this paper, we analyze the macrodiverity gain in the presence of correlated random blocking and interference. To do so, we develop a framework to determine distributions for the LOS probability, SNR, and SINR by taking into account correlated blocking. We consider a cellular uplink with both diversity combining and selection combining schemes. We also study the impact of blockage size and blockage density. We show that blocking can be both a blessing and a curse. On the one hand, the signal from the source transmitter could be blocked, and on the other hand, interfering signals tend to also be blocked, which leads to a completely different effect on macrodiversity gains. We also show that the assumption of independent blocking can lead to an incorrect evaluation of macrodiversity gain, as the correlation tends to decrease macrodiversity gain.
\end{abstract}


\vspace{-0.3cm}

\section{Introduction}

Millimeter-wave (mmWave) has emerged in recent years as a viable candidate for infrastructure-based (i.e., cellular) systems \cite{Rapp:2013,Akdeniz,Andrews17}. At mmWave frequencies, signals are prone to blocking by objects intersecting the paths and severely reducing the signal strength, and thus the Signal to Noise Ratio (SNR) \cite{rappaport2014millimeter}. On the battlefield, blockages may include soldiers, tanks, helicopters, and other equipment creating a dynamic environment characterized by changing blocking conditions. Blocking makes it especially difficult to provide universal coverage with a cellular infrastructure. For instance, blocking by walls provides isolation between indoor and outdoor environments, making it difficult for an outdoor base station to provide coverage indoors \cite{BaiVazeHeath}. To mitigate the issue of blocking in mmWave cellular networks,  macrodiversity has emerged as a promising solution, where the user attempts to connect to multiple base stations. With macrodiversity, the probability of having at least one line-of-sight (LOS) path to a base station increases, which can improve the system performance \cite{Heath2019,Gupta19}. 

An effective methodology to study wireless systems in general, and mmWave systems in particular, is to embrace the tools of stochastic geometry to analyze the SNR and interference in the network \cite{Venugopal2016,Hriba2017,Enass2018,Andrews17}. With stochastic geometry, the locations of base stations and blockages are assumed to be drawn from an appropriate point process, such as a Poisson point process (PPP). When blocking is modeled as a random process, the probability that a link is LOS is an exponentially decaying function of link distance. While many papers assume that blocking is independent \cite{BaiVazeHeath}, in reality the blocking of multiple paths may be correlated \cite{Enass2018}. The correlation effects are especially important for macrodiverity networks when base stations are close to each other, or more generally when base stations have a similar angle to the transmitter. In this case, when one base station is blocked, there is a significant probability that another base station is also blocked \cite{Heath2019,Gupta19}.

Correlated blocking has previously been considered in \cite{Aditya17,Selvatori2018} for localization applications and in \cite{Enass2018} for wireless personal area networks.  Correlation has been considered in \cite{Heath2019,Gupta19} for infrastructure-based networks with macrodiversity, but in these references the only performance metric considered is the $n^{th}$ order LOS probability; i.e., the probability that at least one of the n closest base stations is LOS.  However, a full characterization of performance requires other important performance metrics, including the distributions of the SNR and, when there is interference, the Signal to Interference and Noise Ratio (SINR).  In this paper, we propose a novel approach for fully characterizing the performance of macrodiversity in the presence of correlated blocking.  While, like \cite{Heath2019,Gupta19}, we are able to characterize the spatially averaged LOS probability (i.e., the LOS probability averaged over many network realizations), our analysis shows the \emph{distribution} of the LOS probability, which is the fraction of network realizations that can guarantee a threshold LOS probability rather than its mere spatial average.  Moreover, we are able to similarly capture the distributions of the SNR and SINR.


We assume that the centers of the blockages are placed according to a PPP. We first analyze the distributions of LOS probability for first- and second-order macrodiversity. We then consider the distribution of SNR and SINR for the cellular uplink with both selection combining and diversity combining. The signal model is such that blocked signals are completely attenuated, while LOS, i.e., non-blocked, signals are subject to an exponential path loss and additive white Gaussian noise (AWGN). Though it complicates the exposition and notation, the methodology can be extended to more elaborate models, such as one wherein all signals are subject to fading and non-LOS (NLOS) signals are partially attenuated (see, e.g., \cite{Hriba2017}).

The remainder of the paper is organized as follows. We begin by providing the system model in Section II, wherein there are base stations and blockages, each drawn from a PPP. Section III and section IV provide an analysis of the LOS probability and $\mathsf{SNR}$ distribution, respectively. Then in Section V, interference is considered and the $\mathsf{SINR}$ distribution is formalized. Finally, Section V concludes the paper, suggesting extensions and generalizations of the work.

	\section{System Model}
	

     Consider a mmWave cellular network consisting of base stations, blockages, and a source transmitter. The locations of the base stations are modeled as an infinite homogeneous PPP with density $\lambda_{bs}$. We assume the centers of the blockages also form a homogeneous PPP with density $\lambda_{bl}$, independent from the base station process. Let $Y_0$ indicate the source transmitter and its location. Due to the stationarity of the PPPs, and without loss of generality, we can assume the coordinates are chosen such that the source transmitter is located at the origin; i.e., $Y_0=0$. In section V, we will consider additional transmitters located in neighboring cells, which act as interferers.

		\begin{figure}[t]
		\centering
		\includegraphics[scale=0.4,bb=100 300 150 350]{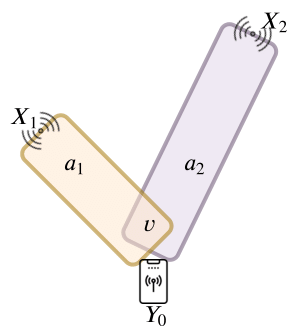}
		\vspace{4cm}
	    \caption{Example network topology for second-order macrodiversity (N=2). $Y_0$ attempts to connect to its closest base stations ($X_1$ and $X_2$). $a_1$ and $a_2$ are the blockage areas, and $v$ is the overlapping area.\vspace{-0.5cm}}
		\label{fig:OverlappingArea}
	\end{figure}

	Let $X_i$ for i $\in {\mathbb Z}^+$ denote the base stations and their locations. Let $R_i = |X_i|$ be the distance from $Y_0$ to $X_i$. Base stations are ordered according to their distances to $Y_0$ such that  $R_1 \leqslant R_2\leqslant...$. The signal of the source transmitter is recived at the closest $N$ base stations, and hence, $N$ is the number of $X_i$ connected to $Y_0$. For a PPP, the derivation of ${R_1, ... ,R_N}$ is given in Appendix A, which implies a methodology for generating these distances within a simulation.
	


	As in \cite{Hriba2017}, each blockage is modeled as a circle of width $W$. Although $W$ can itself be random as in \cite{Heath2019}, we assume here that all blockages have the same value of $W$. If a blockage cuts the path from $Y_0$ to $X_i$, then the signal from $Y_0$ is NLOS, while otherwise it is LOS.  Here, we assume that NLOS signals are completely blocked while LOS signals experience exponential path-loss with a path-loss exponent $\alpha$; i.e., the power received by $X_i$ is proportional to $R_i^{-\alpha}$.

Each base station has a blockage region associated with it, indicated by the colored rectangles shown in Fig. \ref{fig:OverlappingArea}. We use $a_i$ to denote the blockage region associated with $X_i$ and its area; i.e., $a_i$ is both a region and an area. If the center of a blockage falls within $a_i$, then $X_i$ will be blocked since at least some part of the blockage will intersect the path between $X_i$ and $Y_0$. Because $a_i$ is a rectangle of length $R_i$ and width $W$, it is clear that $a_i = W R_i$. Unless $X_1$ and $X_2$ are exactly on opposite sides of the region, there will be an overlapping region $v$ common to both $a_1$ and $a_2$. Because of the overlap, it is possible for a single blockage to simultaneously block both $X_1$ and $X_2$ if the blockage falls within region $v$, which corresponds to correlated blocking.

   \section{LOS probability} 
	In this section, we will analyze the LOS probability ($p_{LOS}$) and its distribution. With second-order macrodiversity, the signal of the source transmitter $Y_0$ is received at the two closest base stations $X_1$ and $X_2$. The LOS probability is the probability that at least one $X_i$ is LOS to the transmitter. Because the base stations are randomly located, the value of $p_{LOS}$ will vary from one network realization to the next, or equivalently by a change of coordinates, from one source transmitter location to the next. Hence $p_{LOS}$ is itself a random variable and must be described by a distribution. To determine $p_{LOS}$ and its distribution, we first need to define the variable $B_i$ which indicates that the path between $Y_0$ and $X_i$ is blocked.  Let $p_{B_{1},B_{2}}(b_{1},b_{2})$ be the joint probability mass function (pmf) of $\{ B_{1}, B_{2} \}$. Let $p_{i}$ denote the probability that $B_{i} = 1$, which indicates the link from $Y_0$ to $X_i$ is NLOS.  Furthermore, let $q_{i} = 1 - p_{i}$, which is the probability that the link is LOS, and $\rho$ denote the correlation coefficient between $B_{1}$ and $B_{2}$. The joint pmf of $\{ B_{1}, B_{2} \}$ as a function of $\rho$ can be found in Appendix B.

Because blockages are drawn from a PPP with density $\lambda_{bl}$, the probability that at least one blockage lands in $a_{i}$ is $p_i=1-\exp(-\lambda_{bl} a_{i})$, and when this occurs $X_i$ will be NLOS. Conversely, $X_i$ will be LOS when there are no blockages inside $a_{i}$, which occurs with probability $q_i= \exp(-\lambda_{bl} a_{i})$. For second-order macrodiversity, there will be a LOS signal as long as both paths are not blocked.  Considering independent blocking, the corresponding LOS probability is $1-p_{1}p_{2}$. However from (\ref{eq:pB1B2}) in Appendix B, when correlated blocking is considered, $p_{LOS}$ is
	\begin{eqnarray}
	p_{LOS}=1 - p_{B_1,B_2}(1,1) =1-p_{1}p_{2}-\rho h \label{eq:pi}
	\end{eqnarray}
where $h = \sqrt{p_1 p_2 q_1 q_2}$, and $\rho$ is found from (\ref{eq:pB1B2}) as	 
\begin{eqnarray}
  \rho
  & = &
  \frac{ p_{B_{1},B_{2}}(0,0) - q_{1} q_{2} } {h}
  \label{eq:rho}
\end{eqnarray}
where $p_{B_{1},B_{2}}(0,0)$ is the probability that both $X_1$ and $X_2$ are not blocked. Looking at Fig. \ref{fig:OverlappingArea}, this can occur when there are no blockages inside $a_{1}$ and $a_{2}$. Taking into account the overlap $v$, this probability is 
\vspace{-0.1cm}
\begin{eqnarray}
 p_{B_{1},B_{2}}(0,0)= e^{-\lambda_{bl}(a_{1}+a_{2}-v)}
 \label{eq:p00}
\end{eqnarray}
Details on how to compute the overlapping area $v$ are provided in \cite{Enass2018}. Substituting (\ref{eq:p00}) into (\ref{eq:rho}) into (\ref{eq:pi}) and using the definitions of $p_i$ and $q_i$, yields
\begin{eqnarray}
p_{LOS}=e^{-\lambda_{bl} a_{1}}+e^{-\lambda_{bl} a_{2}}-e^{-\lambda_{bl}(a_{1}+a_{2}-v)}
 \label{eq:pLOS}
\end{eqnarray}%
	\begin{figure}[t]
		\centering
		\includegraphics[width=1.1\textwidth]{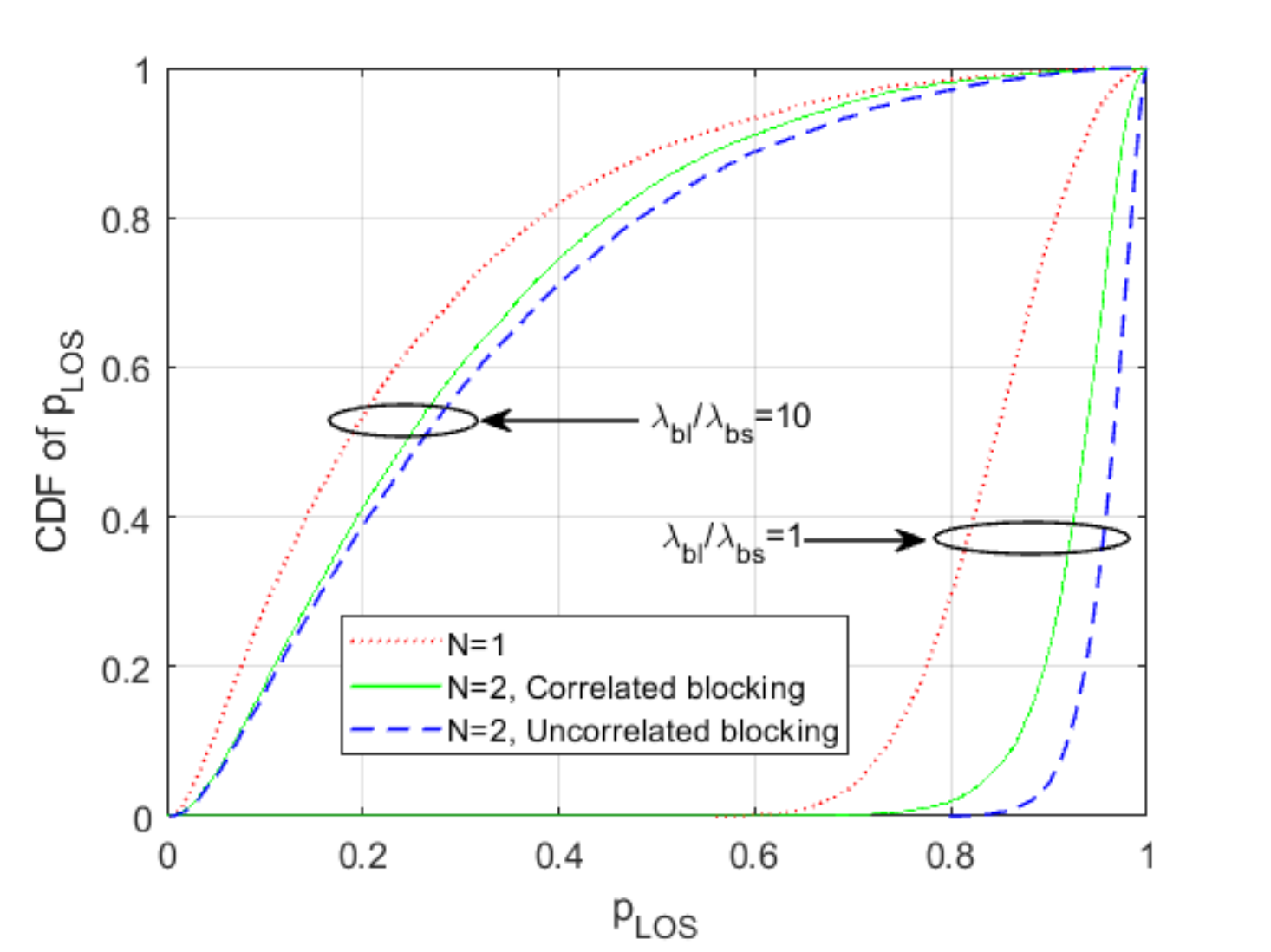}
		\caption{The empirical CDF of $p_{LOS}$ over 1000 network realizations when $N={1,2}$, with and without considering blockage correlation at fixed blockage width $W=0.8$.\vspace{-0.1cm}}
		\label{fig:cdf_SNR_lambda_bl_rate}
	\end{figure}	
	\begin{figure}[t]
	\centering
	\includegraphics[width=\textwidth]{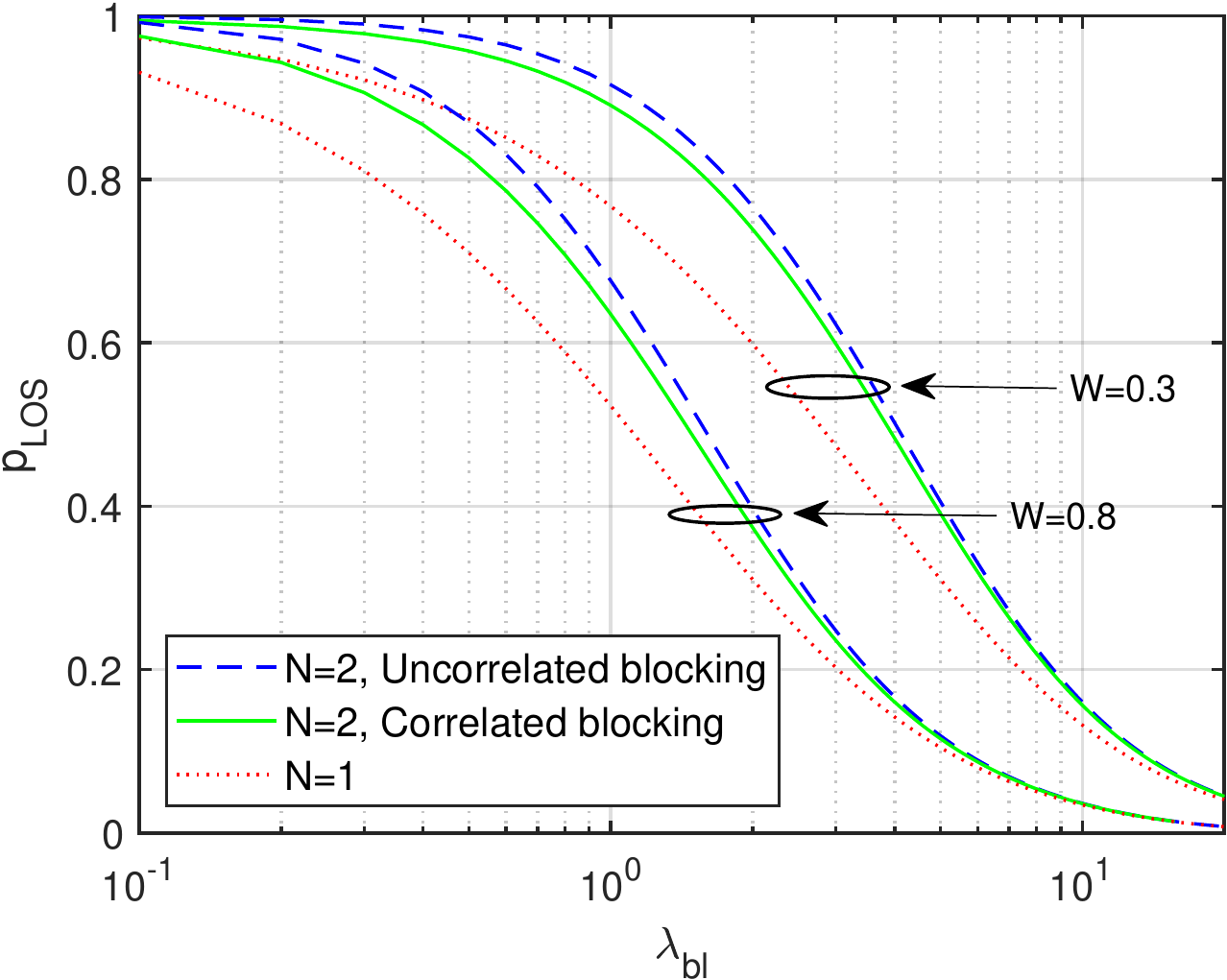}
	\caption{The variation of the spatially averaged $p_{LOS}$ over 1000 network realizations with respect to blockage density $\lambda_{bl}$ when $N={1,2}$, with and without considering blockage correlation at fixed base station density $\lambda_{bs}=0.3$.\vspace{-0.5cm}}
	\label{fig:plos_vs_lambdabl}
\end{figure}		

Fig. \ref{fig:cdf_SNR_lambda_bl_rate} shows the empirical CDF of $p_{LOS}$ over 1000 network realizations for first- and second-order macrodiversity, both with and without considering blockage correlation. The distributions are computed for two different values of the average number of blockages per base station ($\lambda_{bl}/\lambda_{bs}$). We fix the value of $W$ at $0.8$. The CDF of $p_{LOS}$ quantifies the likelihood that the $p_{LOS}$ is below some value. The figure shows the probability that $p_{LOS}$ is below some value increases significantly when the number of blockages per base station is high. The effect of correlated blocking is more pronounced when there are fewer blockages per base station.  Compared to when $N=1$, the macrodiversity gain is higher when the number of blockages per base station is lower even though the amount of reduction in gain due to correlation is higher when $\lambda_{bl}/\lambda_{bs}$ is lower.

Fig. \ref{fig:plos_vs_lambdabl} shows the variation of $p_{LOS}$ when averaged over 1000 network realizations.  The plot shows average $p_{LOS}$ as a function of blockage density $\lambda_{bl}$ while keeping base station density $\lambda_{bs}$ fixed at 0.3. The spatially averaged $p_{LOS}$ is computed for two different values of blockage width $W$. Compared to the case of no diversity (when $N=1$), the second-order macrodiversity can significantly increase $p_{LOS}$. However, $p_{LOS}$ decreases when blockage size or blockage density is higher. Moreover, larger blockages increase the correlation. This is because larger blockages can increase the correlation and decreases macrodiversity gain, since a single large blockage is likely to simultaneously block both base stations. Comparing the two pairs of correlated/uncorrelated blocking curves, the correlation is more dramatic when $\lambda_{bl}$ is low, since at low $\lambda_{bl}$ both base stations are typically blocked by the same blockage (located in area $v$).
\vspace{-0.08cm}  
\section{SNR Distribution}
In this section, we consider the distribution of the $\mathsf{SNR}$.  Macrodiversity can be achieved by using either diversity combining, where the signals from the multiple base stations are maximum ratio combined, or selection combining, where only the signal with the strongest $\mathsf{SNR}$ is used. For second-order macrodiversity, the $\mathsf{SNR}$ with diversity combining is
	\begin{eqnarray}
	\mathsf{SNR} = \mathsf{SNR_0}\underbrace{\sum_{i=1}^2 (1-B_{i})\Omega_{i}}_{Z}
	\label{eg:S}
	\end{eqnarray}
	where ${\Omega_{i}}=R_{i}^{-\alpha}$ is the power gain between the source transmitter $Y_0$ to the $i^\text{th}$ base station and $\mathsf{SNR_0}$ is the $\mathsf{SNR}$ of an unblocked reference link of unit distance. $B_{i}$ is used to indicate that the path between $Y_0$ and $X_i$ is blocked, and thus when $B_{i} = 1$, $\Omega_{i}$ does not factor into the $\mathsf{SNR}$. 
	The CDF of $\mathsf{SNR}$, $F_\mathsf{SNR}(\beta)$, quantifies the likelihood that the combined $\mathsf{SNR}$ at the closest two base stations is below some threshold $\beta$.  If $\beta$ is interpreted as the minimum acceptable $\mathsf{SNR}$ required to achieve reliable communications, then $F_\mathsf{SNR}(\beta)$ is the \emph{outage probability} of the system $P_o(\beta) = F_\mathsf{SNR}(\beta)$.  The \emph{coverage probability} is the \emph{complimentary} CDF, $P_c(\beta) = 1-F_\mathsf{SNR}(\beta)$ and is the likelihood that the $\mathsf{SNR}$ is sufficiently high to provide coverage.  The \emph{rate} distribution can be found by linking the threshold $\beta$ to the transmission rate, for instance by using the appropriate expression for channel capacity.
	
	The CDF of $\mathsf{SNR}$ evaluated at threshold $\beta$ can be determined as follows:\vspace{-0.2cm}
	\begin{eqnarray}
	F_\mathsf{SNR}(\beta)&=&P\left[\mathsf{SNR}\leq \beta \right]\nonumber \\
	&=&P\bigg[Z \leq \frac{\beta}{\mathsf{SNR_0}}\bigg]\nonumber \\
	&=&F_{Z}\left(\frac{\beta}{\mathsf{SNR_0}}\right).\label{cdfsinr}
	\end{eqnarray}

 To find the CDF of $Z$ we need to find the probability of each value of $Z$, which is found as follows.
	The probability that $Z=0$ can be found by noting that $Z=0$ when both $X_1$ and $X_2$ are blocked. From (\ref{eq:pB1B2}), this  is
	\begin{eqnarray}
	p_Z(0)=p_{B_{1},B_{2}}(1,1)=p_{1}p_{2}+\rho h. \label{eq:Z=0}
	\end{eqnarray}
	The probability that $Z=\Omega_{i} , i \in \{1,2\}$ can be found by noting that $Z=\Omega_{i}$ when only $X_i$ is LOS. From (\ref{eq:pB1B2}), this  is
	\begin{eqnarray}
	p_Z(\Omega_{1})&=&p_{B_{1},B_{2}}(0,1)=q_{1} p_{2}-\rho h. \label{eq:Z=1} \\
	p_Z(\Omega_{2})&=&p_{B_{1},B_{2}}(1,0)=p_{1} q_{2}-\rho h. \label{eq:Z=2}
	\end{eqnarray}
	Finally, by noting that $Z=\Omega_{1}+\Omega_{2}$ when both $X_1$ and $X_2$ are LOS leads to
	\begin{eqnarray}
	p_Z({\Omega_{1}+\Omega_{2}})&=&p_{B_{1},B_{2}}(0,0)=q_{1} q_{2}+\rho h. \label{eq:Z=1+2}
	\end{eqnarray}
	%
	From (\ref{eq:Z=0}) to (\ref{eq:Z=1+2}), the CDF of $Z$ is found to be:
	\begin{eqnarray}
	F_{Z}\hspace{-0.1cm}\left(z\right)\hspace{-0.1cm}=\hspace{-0.1cm}\begin{cases}
	0       & \smallskip \hspace{-0.2cm} \text{for} \enskip  z<0 \\
	p_{1}p_{2} + \rho h & \smallskip \hspace{-0.2cm}\text{for} \enskip  0\leq z <\Omega_2 \\
	p_{1}  & \smallskip \hspace{-0.2cm}\text{for} \enskip \Omega_2 \leq z<\Omega_1 \\
	p_{1} +q_{1}p_{2}-\rho h & \smallskip \hspace{-0.2cm}\text{for} \enskip \Omega_1 \leq z<  \Omega_1+\Omega_2  \\  
	1 & \smallskip \hspace{-0.2cm}\text{for} \enskip z \geq  \Omega_1+\Omega_2.  
	\end{cases}\hspace{-.3cm}
	\end{eqnarray}
Next, in the case of selection combining, the $\mathsf{SNR}$ is
	\begin{eqnarray}
\mathsf{SNR} = \mathsf{SNR_0} \;  \underbrace{\mathsf{max}\Bigg[{(1-B_{1})\Omega_{1}} 
, {(1-B_{2})\Omega_{2}} \Bigg]}_Z
\label{eg:SNR_n}
\end{eqnarray}
and its CDF, from (\ref{eq:Z=0}) to (\ref{eq:Z=2}) is found to be:	
	\begin{eqnarray}
	F_{Z}\left(z\right)=\begin{cases}
	0       & \smallskip  \text{for} \enskip  z<0 \\
	p_1p_2 + \rho h & \smallskip \text{for} \enskip  0\leq z <\Omega_2 \\
	p_1 & \smallskip \text{for} \enskip \Omega_2 \leq z<  \Omega_1  \\  
	1 & \smallskip \text{for} \enskip z \geq  \Omega_1.
	\end{cases}
	\label{eq:CDF}
	\end{eqnarray}
\begin{figure}[t]
	\centering
	\includegraphics[width=\textwidth]{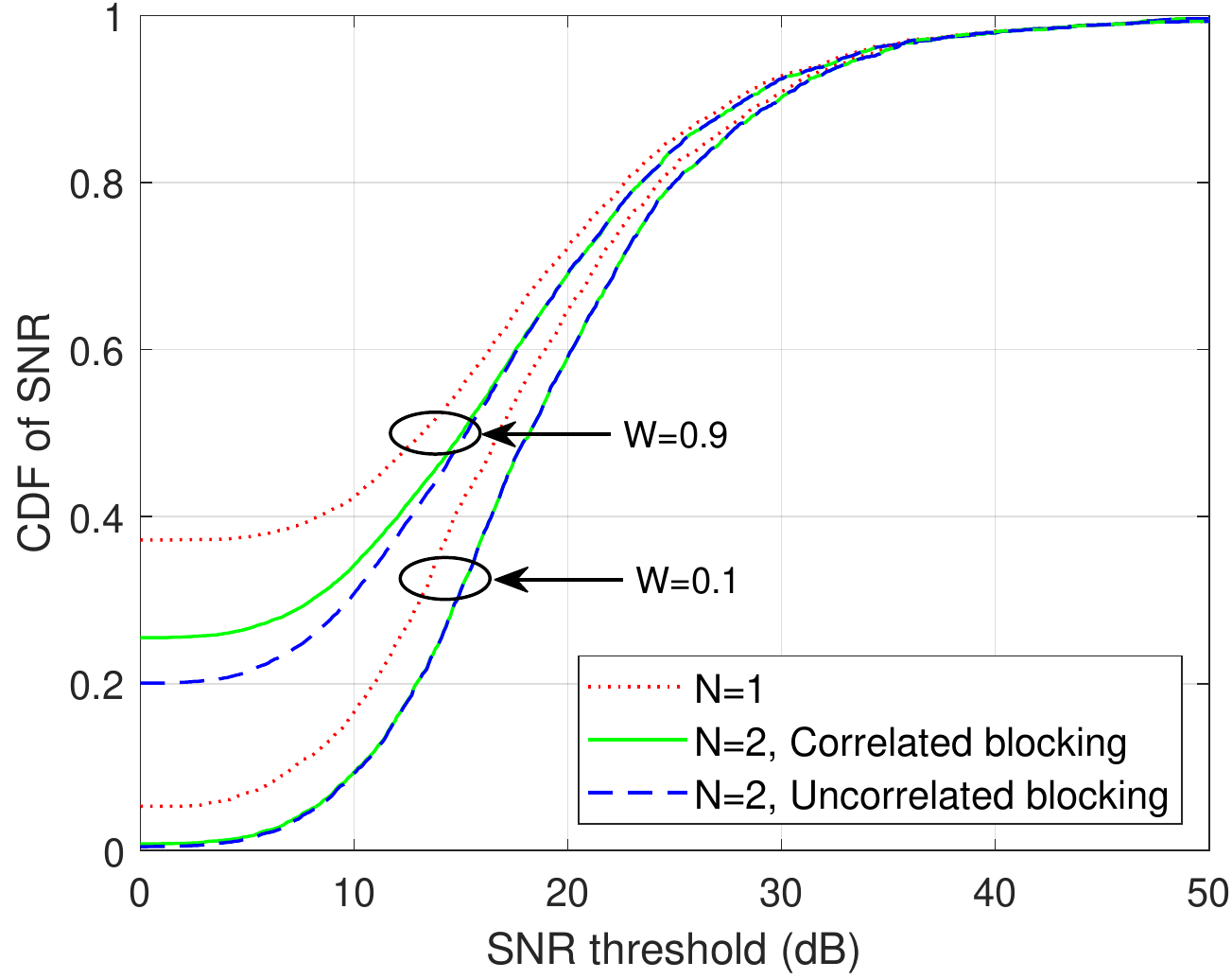}
	\caption{ The distribution of $\mathsf{SNR}$ over 1000 network realizations when $N={1,2}$ using diversity combining, with and without considering blockage correlation at fixed values of blockage density $\lambda_{bl}=0.6$ and base station density $\lambda_{bs}=0.3$.}
	\label{fig:SNR_Diversitylll2}
\end{figure}
\begin{figure}[t]
		\centering
		\includegraphics[width=\textwidth]{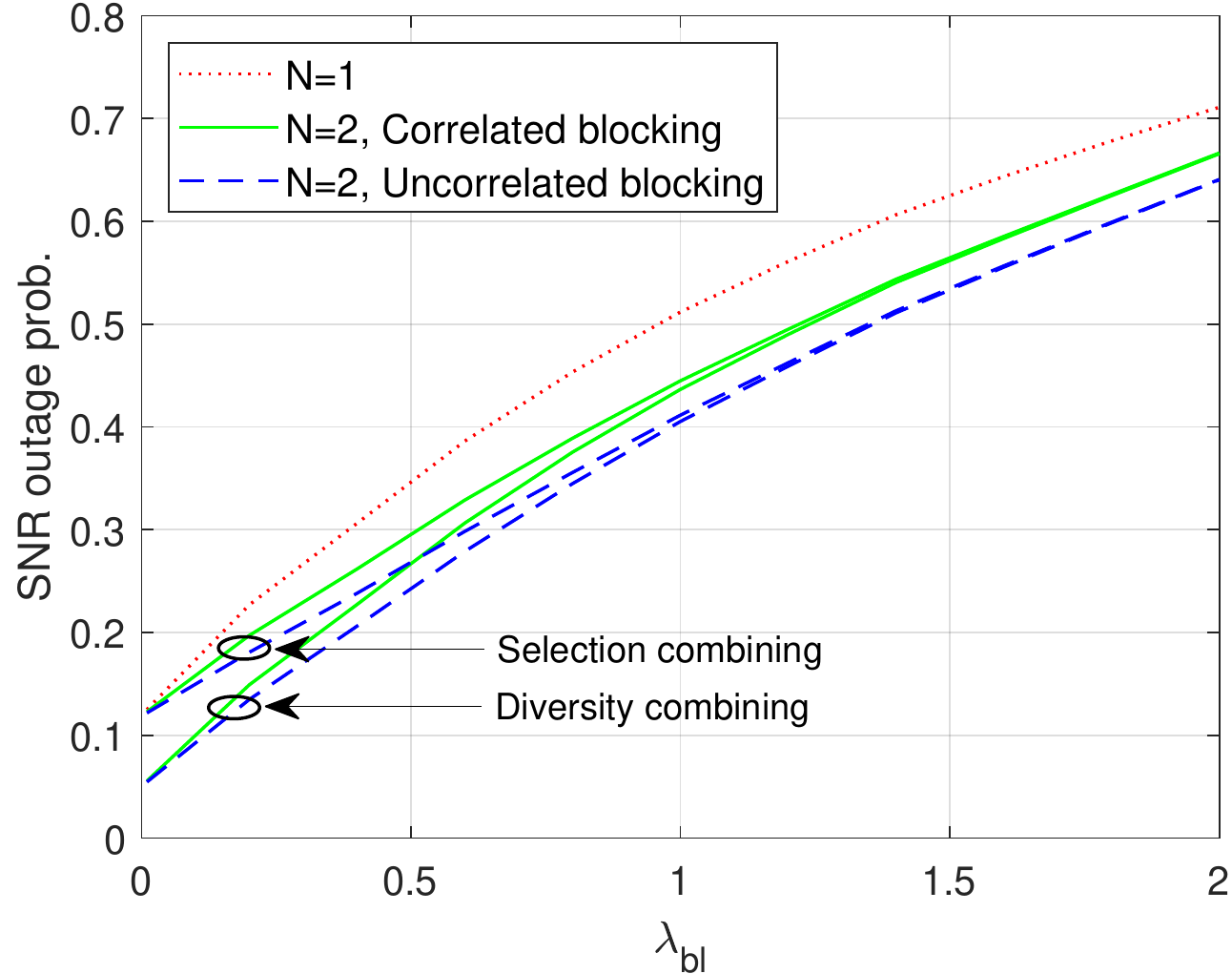}
		\caption{The $\mathsf{SNR}$ outage probability at threshold $\beta=10$ dB with respect to $\lambda_{bl}$ when $N={1,2}$, with and without considering blockage correlation at fixed values of blockage density $\lambda_{bs}=0.3$ and blockage width $W=0.8$.} 
		\label{fig:cdf_SNR_lambda_blL_rate2}
	\end{figure}
 Fig. \ref{fig:SNR_Diversitylll2} shows the CDF of $\mathsf{SNR}$ over 1000 network realizations for diversity combining and two different values of $W$. We fix  $\lambda_{bs}$ at $0.3$ and $\lambda_{bl}$ at $0.6$. In addition, $\mathsf{SNR_0}$ and the path loss $\alpha$ are fixed at $15$ dB and $3$ respectively for the remaining figures in this paper. It can be observed that the distribution increases when blockage size is bigger. Compared to the case when $N=1$, the use of second-order macrodiversity decreases the $\mathsf{SNR}$ distribution. When compared to uncorrelated blocking curves, correlation decreases the gain of macrodiversity for certain regions of the plot, particularly at low values of $\mathsf{SNR}$ threshold, corresponding to the case when both base stations are blocked. Similar to $p_{LOS}$, the correlation increases with blockage size. However, the macrodiversity gain is slightly higher when blockage width $W$ is smaller.

   Fig. \ref{fig:cdf_SNR_lambda_blL_rate2} shows the effect of combining scheme and $\lambda_{bl}$ on $\mathsf{SNR}$ outage probability at threshold $\beta=10$ dB. As shown in the figure, the outage probability increases when $\lambda_{bl}$ increases in all of the given scenarios. When $\lambda_{bl}=0$, first- and second-order selection combining perform identically. This is because $X_1$ is never blocked. However, as $\lambda_{bl}$ increases, the gain of both selection combining and diversity combining increase. At high $\lambda_{bl}$ the combining scheme is less important, in which case the paths to $X_1$ and $X_2$ are always blocked regardless of the chosen combining scheme. The reduction in gain due to correlation is slightly higher when using selection combining. From eq. (\ref{eq:CDF}) this is because the step when both base stations are blocked is wider compared to diversity combining case.

\section{SINR Distribution}

Thus far, we have not assumed any interfering transmitters in the system. In practice, the received signal is also affected by the sum interference. In this section, we assume each neighboring cell has a single interfering mobile, which is located uniformly within a disk of radius $r$ around the base station. Assuming a perfect packing of cells, $r=(\lambda_{bs}\pi)^{-1/2}$, which is the average cell radius. We explicitly consider the interference from the $M$ closest neighboring cells. The interference from more distant cells is considered to be part of the thermal noise.  Let $Y_j$ for $j=1,2,..,M$ indicate the interfering transmitters and their locations. Recall that $j=0$ indicates the source transmitter $Y_0$. The distance from the $j^{th}$ transmitter to the $i^{th}$ base station is denoted by $R_{i,j}$.

To calculate $\mathsf{SINR}$ and its distribution, we first  define a matrix $\mathbf{B}$ which indicates the blocking state of the paths from $Y_j$ for $j=0,2,..,M$ to $X_i$ for $i={1,2}$. $\mathbf{B}$ is a Bernoulli Matrix of size $2$ by $(M+1)$ elements. Each column in $\mathbf{B}$ contain elements $B_{1,j}$ and $B_{2,j}$ which indicate the blocking states of the paths from $Y_j$ to $X_1$ and $X_2$ respectively; i.e, the first column in $\mathbf{B}$ contains the pair of Bernoulli random variables $B_{1,0}$ and $B_{2,0}$ that indicates the blocking state of the paths from $Y_0$ to $X_i$ for $i={1,2}$. There are $(M+1)$ pairs of Bernoulli random variables, and each pair is correlated with correlation coefficient $\rho_j$. Because the $2(M+1)$ elements of $\mathbf{B}$ are binary, there are $2^{2(M+1)}$ possible combinations of $\mathbf{B}$. However, it is possible for different realizations of $\mathbf{B}$ to correspond to the same value of $\mathsf{SINR}$. For example, when $X_1$ and $X_2$ are both blocked from $Y_0$, the $\mathsf{SINR}$ will be the same value regardless of the blocking states of the interfering transmitters. Define $\mathbf{B}^{(n)}$ for $n=1,2,...,2^{2(M+1)}$ to be the $n^{th}$ such combination of $\mathbf{B}$. Similar to Section III, let $p_{B_{1,j},B_{2,j}}(b_{1,j}^{(n)},b_{2,j}^{(n)})$ be the joint probability of $B_{1,j}$ and $B_{2,j}$ which are the elements of the $j^{th}$ column of $\mathbf{B}^{(n)}$. The probability of $\mathbf{B}^{(n)}$ is given by  \vspace{-0.15cm}
\begin{eqnarray}
P(B^{(n)})=\prod_{j=0}^M p_{B_{1,j},B_{2,j}}(b_{1,j}^{(n)},b_{2,j}^{(n)})
\end{eqnarray}
The $\mathsf{SINR}$ of a given realization $\mathbf{B}^{(n)}$ at base station $X_i$ is given by  \vspace{-0.29cm}
\begin{eqnarray}
\mathsf{SINR_i^{(n)}} = \frac{(1-B_{i,0}^{(n)})\Omega_{i,0}}{\mathsf{SNR}_0^{-1}+\displaystyle \sum_{j=1}^{M} (1-B_{i,j}^{(n)})\Omega_{i,j}}
\label{eg:SINRi}
\end{eqnarray}
where $\Omega_{i,j}=R_{i,j}^{-\alpha}$ is the path gain from the $j^{th}$ transmitter at the $i^{th}$ base station. The $\mathsf{SINR}$ of the combined signal considering selective combining is expressed as\vspace{-0.1cm}
\begin{eqnarray}
\mathsf{SINR^{(n)}}= \mathsf{max} \left( \mathsf{SINR_1^{(n)}},\mathsf{SINR_2^{(n)}}\right)
\label{eg:SINRb_n2}
\end{eqnarray}
When considering diversity combining (\ref{eg:SINRb_n2}) changes to
\begin{eqnarray}
\mathsf{SINR^{(n)}}\leq\mathsf{SINR_1^{(n)}}+\mathsf{SINR_2^{(n)}}
\label{eg:SINR_n}
\end{eqnarray}
As described in \cite{int2019}, correlated interference tends to make the combined $\mathsf{SINR}$ less than the sum of the individual $\mathsf{SINRs}$. The bound in (\ref{eg:SINR_n}) is satisfied with equality when the interference is independent at the two base stations.

  To generalize the formula for any realization, there is a particular $\mathsf{SINR^{(n)}}$ associated with each $\mathbf{B}^{(n)}$. However, as referenced above, multiple realizations of $\mathbf{B}^{(n)}$ may result in the same $\mathsf{SINR}$. Let $\mathsf{SINR^{(k)}}$ be the $k^{th}$ realization of $\mathsf{SINR}$. Its probability is \vspace{-0.1cm}
\begin{eqnarray}
P\left(\mathsf{SINR}^{(k)}\right)=\sum_{\substack{
   n:\mathsf{SINR}=\mathsf{SINR}^{(k)}
  }} P\left(B^{(n)}\right)
 \end{eqnarray} 
\begin{figure}[t]
	\centering
	\includegraphics[width=\textwidth]{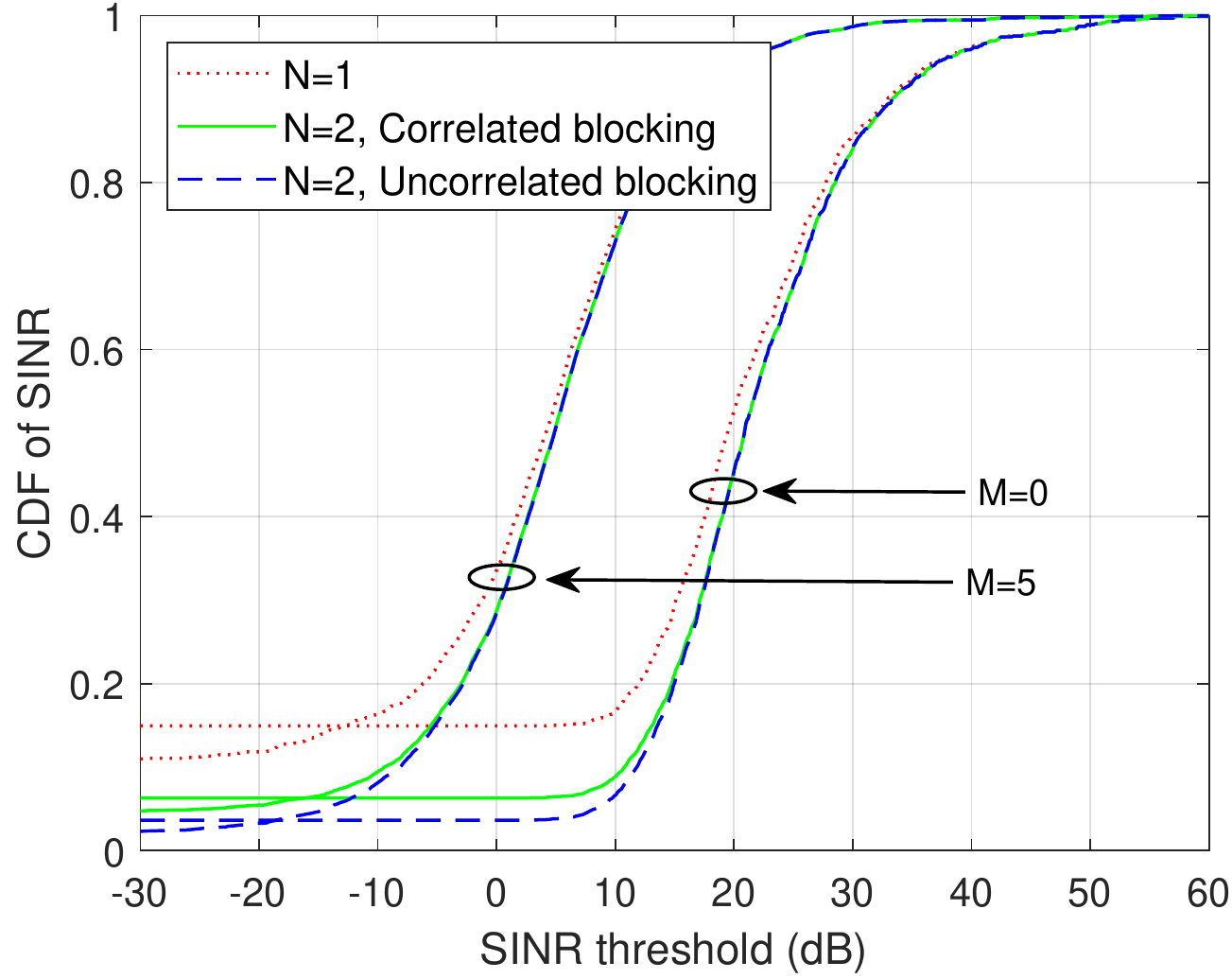}
	\caption{The distribution of $\mathsf{SINR}$ over 1000 network realizations using diversity combining for different values of number of interfering transmitters. The curves are computed when $N={1,2}$, with and without considering blockage correlation, at fixed values of $\lambda_{bs}=0.3$, $\lambda_{bl}=0.6$, and $W=0.6$.\vspace{-0.2cm}}
	\label{fig:CDFSINR_vs_Minter}
\end{figure}

\begin{figure}[t]
	\centering
	\includegraphics[width=\textwidth]{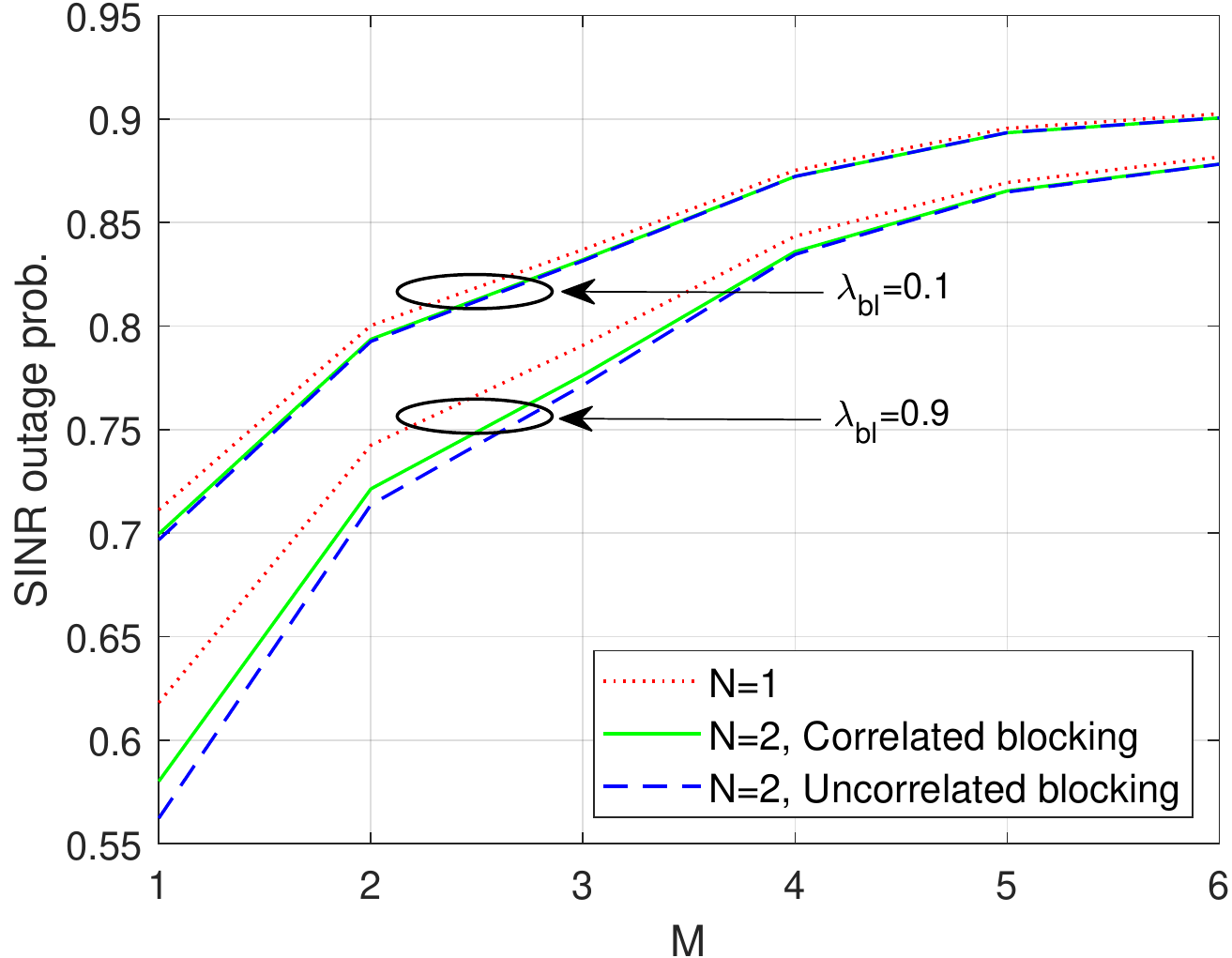}
	\caption{ The outage probability of $\mathsf{SINR}$ at threshold $\beta=15$ dB versus the number of interfering transmitters $(M)$, when $N={1,2}$, with and without considering blockage correlation, at fixed values of $\lambda_{bs}=0.8$ and $W=0.6$.\vspace{-0.6cm}}
	\label{fig:SINRR_vs_diffM}
\end{figure}

Fig. \ref{fig:CDFSINR_vs_Minter} shows the distributions of $\mathsf{SINR}$ for $M=5$ and $M=0$ (which is $\mathsf{SNR}$) at fixed values of $\lambda_{bs}=0.3$, $\lambda_{bl}=0.6$, and $W=0.6$. The distributions are computed for first- and second-order macrodiversity. It can be observed that macrodiversity gain is reduced when interference is considered. This is because of the increase in sum interference due to macrodiversity, which implies that $p_{LOS}$ alone as in \cite{Heath2019} may not be sufficient to predict the performance of the system especially when there are many interfering transmitters. Study of higher order macrodiversity to identify the minimum order of macrodiversity to achieve a desired level of performance in the presence of interference is left for future work.  

Fig. \ref{fig:SINRR_vs_diffM} shows the variation of $\mathsf{SINR}$ outage probability with respect to the number of interfering transmitters $M$. The curves are computed for low and high values of $\lambda_{bl}$, while keeping $\lambda_{bs}$ and $W$ fixed at $0.8$ and $0.6$ respectively. It can be seen that the outage probability increases when $M$ increases. Due to the fact that interference tends to also be blocked, unlike $\mathsf{SNR}$ and $p_{LOS}$, increasing the $\lambda_{bl}$ decreases the outage probability. Similar to Fig.\ref{fig:CDFSINR_vs_Minter}, the macrodiversity gain decreases significantly when $M$ increases. It can be seen that $N=2$ curves reaches the case when $N=1$ for $M=6$. Compared to uncorrelated blocking, the curves considering correlated blocking reaches the uncorrelated cases for high value of $M$, since the interfering transmitters are placed farther than source transmitter and their overlapping area is less dominant. 


\vspace{-0.09cm}
\section{Conclusion}

We have proposed a framework to analyze the second-order macrodiversity gain for mmWave cellular system in the presence of correlated blocking. Correlation is an important consideration for macrodiversity because a single blockage can block multiple base stations, especially if the blockage is sufficiently large and the base stations sufficiently close. The assumption of independent blocking leads to an incorrect evaluation of macrodiversity gain of the system. By using the methodology in this paper, the correlation between two base stations is found and factored into the analysis. The paper considered the distributions of LOS probability, SNR, and, when there is interference, the SINR. We show that correlated blocking decreases the macrodiversity gain. We also study the impact of blockage size and blockage density. We show that blockage can be both a blessing and a curse. On the one hand, the signal from the source transmitter could be blocked, and on the other hand, interfering signals tend to also be blocked, which leads to a completely different effect on macrodiversity gains.

The analysis can be extended in a variety of ways. In Section IV, we have already shown that any number of interfering transmitters can be taken in to account.  While this paper has focused on the extreme case that LOS signals are AWGN while NLOS signals are completely blocked, it is possible to adapt the analysis to more sophisticated channels, such as those where both LOS and NLOS signals are subject to fading and path loss, but the fading and path loss parameters are different depending on the blocking state.  See, for instance, \cite{Hriba2017} for more detail. We may also consider the use of directional antennas, which will control the effect of interference \cite{Torrieri2016}. 

Finally, while this paper focused on second-order macrodiversity, the study can be extended to the more general case of an arbitrary macrodiversity order. Such a study could identify the minimum macrodiversity order required to achieve desired performance in the presence of interference. We anticipate that when more than two base stations are connected, the effects of correlation on macrodiversity gain will increase and the effect of interference will decrease. This is because the likelihood that two base stations are close together increases with the number of base stations and the ratio of the number of connected base stations to the number of interfering transmitters will increase.


\section*{Appendix A}
From the pdf of $R_i$ given in \cite{Heath2019}, we can derive the CDF of $R_i$ given $R_{i-1}$ as
\begin{eqnarray}
F_{R_i}(r_i|R_{i-1}=r_{i-1})=1-e^{\lambda \pi (r_i^2-r_{i-1}^2)}
	\end{eqnarray}
To generate random variables $r_1, ... , r_N$, let $x_i \sim U(0,1)$,
\begin{eqnarray}
x_i=F_{R_i}(r_i|R_{i-1}=r_{i-1})=1-e^{\lambda \pi (r_i^2-r_{i-1}^2)}
\end{eqnarray}
Solving for $r_i$,
\begin{eqnarray}
r_i=\sqrt{-\frac{1}{\lambda \pi} \ln{(1-x_i)}+r_{i-1}^2}\label{eq:ri}
\end{eqnarray}
where $r_0=0$, Start by generating $x_i$ as uniform random variables, then recursively substitute each one in (\ref{eq:ri}) to get the desired random variable  $r_i.$
%
\section*{Appendix B}
	The joint pmf of $B_1$ and $B_2$ is given by:
	\begin{eqnarray}
	p_{B_1,B_2}(b_1,b_2)
	\hspace{-0.2cm}& = &\hspace{-0.2cm}
	\begin{cases}
	q_1 q_2  +\rho h &  \mbox{for $b_1 = 0, b_2 = 0$} \\  
	q_1 p_2 - \rho h &  \mbox{for $b_1 = 0, b_2 = 1$} \\   
	p_1q_2 - \rho h  &  \mbox{for $b_1 = 1, b_2 = 0$} \\  
	p_1 p_2 + \rho h  &  \mbox{for $b_1 = 1, b_2 = 1$} \\    
	\end{cases}
	\label{eq:pB1B2}
	\end{eqnarray}
	where $h = \sqrt{p_1 p_2 q_1 q_2}$. Proof of (\ref{eq:pB1B2}) can be found in \cite{Enass2018}.

\balance

\small
\bibliographystyle{ieeetr}

\balance
\bibliography{./References}

\begin{thebibliography}{10}

\bibitem{Rapp:2013}
T.~Rappaport~{\em et al}, ``Millimeter wave mobile communications for 5{G}
  cellular: It will work!,'' {\em IEEE Access}, vol.~1, pp.~335--349, 2013.

\bibitem{Akdeniz}
M.~R. Akdeniz, Y.~Liu, M.~K. Samimi, S.~Sun, S.~Rangan, T.~S. Rappaport, and
  E.~Erkip, ``Millimeter wave channel modeling and cellular capacity
  evaluation,'' {\em IEEE Journal on Selected Areas in Communications},
  vol.~32, pp.~1164--1179, June 2014.

\bibitem{Andrews17}
J.~G. Andrews, T.~Bai, M.~N. Kulkarni, A.~Alkhateeb, A.~K. Gupta, and R.~W.
  {Heath, Jr.}, ``Modeling and analyzing millimeter wave cellular systems,''
  {\em IEEE Trans. Commun.}, vol.~65, pp.~403--430, Jan 2017.

\bibitem{rappaport2014millimeter}
T.~Rappaport, R.~W. {Heath, Jr.}, R.~C. Daniels, and J.~N. Murdock, {\em
  Millimeter Wave Wireless Communications}.
\newblock Pearson Education, 2014.

\bibitem{BaiVazeHeath}
T.~Bai, R.~Vaze, and R.~W. {Heath, Jr.}, ``Analysis of blockage effects on
  urban cellular networks,'' {\em IEEE Trans. Wireless Comm.}, vol.~13,
  pp.~5070--5083, Sept. 2014.

\bibitem{Heath2019}
A.~K. Gupta, J.~G. Andrews, and R.~W. Heath, ``Macrodiversity in cellular
  networks with random blockages,'' {\em IEEE Transactions on Wireless
  Communications}, vol.~17, pp.~996--1010, Feb. 2018.

\bibitem{Gupta19}
A.~K. {Gupta}, J.~G. {Andrews}, and R.~W. {Heath}, ``Impact of correlation
  between link blockages on macro-diversity gains in mmwave networks,'' in {\em
  2018 IEEE International Conference on Communications Workshops (ICC
  Workshops)}, May 2018.

\bibitem{Venugopal2016}
K.~Venugopal, M.~C. Valenti, and R.~W. {Heath, Jr.}, ``Device-to-device
  millimeter wave communications: Interference, coverage, rate, and finite
  topologies,'' {\em IEEE Trans. Wireless Comm.}, vol.~15, pp.~6175--6188,
  Sept. 2016.

\bibitem{Hriba2017}
E.~Hriba, M.~C. Valenti, K.~Venugopal, and R.~W. Heath, ``Accurately accounting
  for random blockage in device-to-device {mmWave} networks,'' in {\em Proc.
  IEEE Global Telecommun. Conf. (GLOBECOM)}, Dec. 2017.

\bibitem{Enass2018}
E.~Hriba and M.~C.~Valenti, ``The impact of correlated blocking on
  millimeter-wave personal networks,'' in {\em Proc. IEEE Military Commun.
  Conf. (MILCOM)}, Oct. 2018.

\bibitem{Aditya17}
S.~Aditya, H.~S. Dhillon, A.~F. Molisch, and H.~Behairy, ``Asymptotic
  blind-spot analysis of localization networks under correlated blocking using
  a {Poisson} line process,'' {\em IEEE Wireless Communications Letters},
  vol.~6, pp.~654--657, Oct. 2017.

\bibitem{Selvatori2018}
S.~{Talarico}, M.~C. {Valenti}, and M.~{Di Renzo}, ``Outage correlation in
  finite and clustered wireless networks,'' in {\em 2018 IEEE 29th Annual
  International Symposium on Personal, Indoor and Mobile Radio Communications
  (PIMRC)}, pp.~1--7, Sep. 2018.

\bibitem{int2019}
R.~{Tanbourgi}, H.~S. {Dhillon}, J.~G. {Andrews}, and F.~K. {Jondral}, ``Effect
  of spatial interference correlation on the performance of maximum ratio
  combining,'' {\em IEEE Transactions on Wireless Communications}, vol.~13,
  pp.~3307--3316, June 2014.

\bibitem{Torrieri2016}
D.~Torrieri, S.~Talarico, and M.~C. Valenti, ``Analysis of a frequency-hopping
  millimeter-wave cellular uplink,'' {\em IEEE Trans. Wireless Comm.}, vol.~15,
  pp.~7089--7098, Oct. 2016.

\end{thebibliography}


\end{document}